\documentclass[12pt,preprint]{aastex}

\slugcomment{Submitted to the Astrophysical Journal on October 5, 2005}

\shorttitle{The young, highly relativistic binary pulsar J1906+0746}
\shortauthors{Lorimer et al.}

\begin{document}

\title{Arecibo Pulsar Survey Using ALFA. II. The young, highly
       relativistic binary pulsar J1906+0746}

\author{
  D.~R.~Lorimer,\altaffilmark{1}
  I.~H.~Stairs,\altaffilmark{2}
  P.~C.~C.~Freire,\altaffilmark{3}
  J.~M.~Cordes,\altaffilmark{4}
  F.~Camilo,\altaffilmark{5}
  A.~J.~Faulkner,\altaffilmark{1}
  A.~G.~Lyne,\altaffilmark{1}
  D.~J.~Nice,\altaffilmark{6}
  S.~M.~Ransom,\altaffilmark{7}
  Z.~Arzoumanian,\altaffilmark{8}
  R.~N.~Manchester,\altaffilmark{9}
  D.~J.~Champion,\altaffilmark{1}
  J.~van Leeuwen,\altaffilmark{2}
  M.~A.~McLaughlin,\altaffilmark{1}
  R.~Ramachandran,\altaffilmark{10}
  J.~W.~T.~Hessels,\altaffilmark{11}
  W.~Vlemmings,\altaffilmark{1}
  A.~A.~Deshpande,\altaffilmark{3}
  N.~D.~R.~Bhat,\altaffilmark{12}
  S.~Chatterjee,\altaffilmark{7,13}
  J.~L.~Han,\altaffilmark{14}
  B.~M.~Gaensler,\altaffilmark{13}
  L.~Kasian, \altaffilmark{2}
  J.~S.~Deneva,\altaffilmark{4}
  B.~Reid,\altaffilmark{15}
  T.~J.~W.~Lazio,\altaffilmark{16}
  V.~M.~Kaspi,\altaffilmark{11}
  F.~Crawford,\altaffilmark{17}
  A.~N.~Lommen,\altaffilmark{18}
  D.~C.~Backer,\altaffilmark{10}
  M.~Kramer,\altaffilmark{1}
  B.~W.~Stappers, \altaffilmark{19,20}
  G.~B.~Hobbs,\altaffilmark{9}
  A.~Possenti,\altaffilmark{21}
  N.~D'Amico,\altaffilmark{21}
  and  M.~Burgay\altaffilmark{21}
}
\altaffiltext{1}{University of Manchester, Jodrell Bank Observatory, Cheshire, SK11~9DL, UK}
\altaffiltext{2}{Department of Physics \& Astronomy, University of
British Columbia, 6224 Agricultural Road, Vancouver, B.C. V6T 1Z1, Canada}
\altaffiltext{3}{National Astronomy and Ionosphere Centre, Arecibo Observatory, HC3 Box 53995, PR 00612}
\altaffiltext{4}{Astronomy Department, Cornell University, Ithaca, NY 14853}
\altaffiltext{5}{Columbia Astrophysics Laboratory, Columbia University, 550 West 120th Street, New York, NY~10027}
\altaffiltext{6}{Physics Department, Bryn Mawr College, Bryn Mawr, PA 19010}
\altaffiltext{7}{National Radio Astronomy Observatory, 520 Edgemont Rd, Charlottesville, VA}
\altaffiltext{8}{USRA/EUD, NASA Goddard Space Flight Center, Greenbelt, MD 20771}
\altaffiltext{9}{Australia Telescope National Facility, CSIRO, Epping, NSW 1710, Australia}
\altaffiltext{10}{Astronomy Department, University of California, Berkeley, CA 94720}
\altaffiltext{11}{McGill University Physics Department, Montreal, QC H3A2T8, Canada}
\altaffiltext{12}{Swinburne University of Technology, Centre for Astrophysics and Supercomputing, Victoria 3122, Australia}
\altaffiltext{13}{Harvard-Smithsonian Center for Astrophysics, 60 Garden Street, Cambridge, MA~02138}
\altaffiltext{14}{National Astronomical Observatories, Chinese Academy of
Sciences, A20 DaTun Road, Chaoyang District, Beijing, 100012, China}
\altaffiltext{15}{Physics Department, Princeton University, Princeton, NJ 08544}
\altaffiltext{16}{Naval Research Laboratory, 4555 Overlook Ave. S.W., Washington, DC 20375}
\altaffiltext{17}{Department of Physics, Haverford College, Haverford, PA 19041-1392}
\altaffiltext{18}{Franklin and Marshall College, Department of Physics and Astronomy, Lancaster, PA 17604}
\altaffiltext{19}{Stichting ASTRON, Postbus 2, 7990 AA Dwingeloo, The Netherlands}
\altaffiltext{20}{Astronomical Institute ``Anton Pannekoek'', University of Amsterdam, Kruislaan 403, 1098 SJ Amsterdam, The Netherlands}
\altaffiltext{21}{Istituto Nazionale di Astrofisica, Osservatorio Astronomico di Cagliari, loc. Poggio dei Pini, Strada 54, 09012 Capoterra, Italy}

\begin{abstract}
We report the discovery of PSR~J1906+0746, a young 144-ms pulsar in a highly
relativistic 3.98-hr orbit with an eccentricity of 0.085 and expected
gravitational wave coalescence time of $\sim 300$~Myr. The new pulsar
was found during precursor survey observations with the Arecibo
1.4-GHz feed array system and retrospectively detected in the Parkes
Multibeam plane pulsar survey data. From radio
follow-up observations with Arecibo, Jodrell Bank, Green Bank, and
Parkes, we have measured the spin-down and binary parameters of the
pulsar and its basic spectral and polarization properties. We also
present evidence for pulse profile evolution, which is likely
due to geodetic precession, a relativistic effect caused by 
the  misalignment of the pulsar spin and total
angular momentum vectors. Our measurements show that PSR~J1906+0746 is a
young object with a characteristic age of 112~kyr. From the measured
rate of orbital periastron advance ($7.57\pm0.03$\arcdeg~yr$^{-1}$),
we infer a total system mass of $2.61\pm0.02$~$M_{\odot}$.  While
these parameters suggest that the PSR~J1906+0746 binary system might
be a younger version of the double pulsar system, intensive searches
for radio pulses from the companion have so far been unsuccessful. 
It is therefore not known whether the companion is another neutron star
or a massive white dwarf. Regardless of the nature of the companion,
a simple calculation suggests that the Galactic birth
rate of binaries similar to PSR~J1906+0746 is $\sim 60$~Myr$^{-1}$. This
implies that PSR~J1906+0746 will make a significant contribution to the
computed cosmic inspiral rate of compact binary systems.
\end{abstract}


\section{Introduction}\label{sec:intro}\setcounter{footnote}{0}

Binary radio pulsars in compact short-period orbits around other
neutron stars or white dwarfs are valuable to study for a variety of
reasons. The ability to monitor precisely the pulsars as clocks in
such systems allows exquisite tests of Einstein's theory of general
relativity to be performed (e.g.~Taylor \& Weisberg 1989; Lyne et
al.~2004)\nocite{tw89,lbk+04}. Additionally, studies of the variations
in pulse profile morphology and polarimetry provide probes of
relativistic spin-orbit coupling in several systems \citep[for a
review, see][]{sta03}. As the rare end-points of binary star
evolution, relativistic binaries offer unique insights and powerful
constraints on our physical understanding of the formation properties
of compact objects \citep[e.g.~the natal kicks imparted to the neutron
star;][]{lcc01,wkh04}. Since many of these binaries will coalesce due
to gravitational wave emission well within a Hubble time, their merger
rate \citep{phi91,bdp+03,kkl03,kklw04} is of great interest to the
gravitational wave detector community as potential sources for current
interferometers such as LIGO \citep{aad+92}.

Following the discovery of the original binary pulsar B1913+16
\citep{ht75a}, observational progress in finding further relativistic
binaries was hampered by a lack of detector sensitivity and
computational resources required to detect their strongly time-varying
signals \citep{jk91}. In recent years, as both of these obstacles are
gradually being overcome, the sample of relativistic binaries with
orbital periods less than a day has grown and now numbers eight
\citep{mhth05}.  Here we report the discovery of the latest addition
to this growing population, PSR~J1906+0746, found in the preliminary
stages of a large-scale search for pulsars we are carrying out using
the 305-m Arecibo telescope \citep[][hereafter paper I]{cfl+05}.  This
new binary system promises to have a significant impact on all of the
science areas mentioned above. In \S \ref{sec:discovery} we detail the
discovery and follow-up observations carried out so far. We then
discuss the implications of these results on the nature of the
companion (\S \ref{sec:nature}) and merger rate of compact binary
systems (\S \ref{sec:age}), and present evidence for pulse profile
evolution (\S \ref{sec:profevol}). In \S \ref{sec:future}, we
summarize the main conclusions and look ahead to future work.

\section{Discovery and follow-up observations}\label{sec:discovery}

PSR~J1906+0746, was discovered in precursor observations with the Arecibo
L-band Feed Array\footnote{http://alfa.naic.edu} (ALFA) system
described in paper I. In brief, ALFA collects radio signals from the
Gregorian focus at Arecibo using seven cooled receivers which cover
different parts of the sky in the 1.2--1.5~GHz band. Pulsar survey
observations with the ALFA system (P-ALFA) were carried out in a
precursor phase between 2004 August and 2004 October and covered a
total of 15.8~deg$^2$ in the inner Galaxy ($40^\circ \le l \le
75^\circ$, $|b| \le 1^\circ$) with 135-s pointings and 14.8~deg$^2$ in
the anti-center region ($170^\circ \le l \le 212^\circ$, $|b| \le 1^\circ$) 
with 67-s pointings.  The incoming signals were amplified, filtered
and down-converted before being sampled by four Wideband Arecibo
Pulsar Processors \citep[WAPPs;][]{dsh00}. In the configuration used
for this survey, the WAPPs provided a spectral resolution of
390.625~kHz for both polarizations over a 100-MHz band centered at
1.42 GHz. The polarization pairs were summed and written to disk every
64~$\mu$s.  Initial data reduction was carried out at reduced spectral
frequency and time resolution as described in paper I.  PSR~J1906+0746, was
discovered with a signal-to-noise ratio S/N~$\sim 11$ in data taken on
2004 September 27. The pulsar was 2$\farcm$5 (1.47 beam radii)
from the center of the beam (see Paper I), where the antenna gain is
$\sim$5 times smaller than at its center.

With Galactic coordinates $l=41.6^\circ$ and $b=0.1^\circ$, PSR~J1906+0746
lies in the region of sky covered by the Parkes Multibeam
Pulsar Survey \citep[PMPS; e.g.][]{mlc+01}. Examination of the
search output of the 35-minute PMPS observation
of this position taken on 1998 August 3 showed a 144-ms periodicity
with S/N~$\sim 7$. Although this was below the nominal S/N threshold
of 8--9 of the PMPS, the data were also processed using
techniques designed to retain sensitivity to binary
pulsars. Using the ``stack-and-slide'' algorithm \citep{bc00}
implemented in the PMPS \citep{fsk+04} PSR~J1906+0746 appears with
S/N~$\sim25$. The pulsar was 3\farcm8 (only 0.5 beam radii) from
the center of the beam. However, due to significant amounts of
radio-frequency interference close to 144~ms, PSR~J1906+0746 was not
selected as a candidate in the PMPS.

The high degree of acceleration detected in the PMPS observation immediately
implied that PSR~J1906+0746 is a short-period binary system. To establish
the orbital parameters, follow-up observations were initiated in 2005
May using the 76-m Lovell telescope at Jodrell Bank. The data
acquisition system used for these 1396-MHz observations was identical
to that described by \cite{hlk+04}. A preliminary orbital ephemeris
was obtained from period determinations in 5-min integrations over an
initial 9-hr observation. The ephemeris was then refined in a timing
analysis using the {\sc tempo} software
package\footnote{http://pulsar.princeton.edu/tempo}, where pulse times
of arrival (TOAs) were fitted to a model including spin parameters,
orbital elements and pulsar position.  The resulting ephemeris, based
on 1640 TOAs spanning 158 days, is given in Table~\ref{tab:parms}.  As
a check on the solution, the pulsar position was independently derived
from a grid of Arecibo observations at 3~GHz. The position thus
measured has an uncertainty of 9 arcseconds and is 6.7 arcseconds away
from the timing position.

Although all data used for the timing solution were collected in the
1--2~GHz and 3--4~GHz bands with Arecibo, 
we have also been able to detect PSR~J1906+0746 at 430~MHz
and 6~GHz with Arecibo and at 820~MHz with the Green Bank Telescope
(GBT). The Arecibo data were collected using the WAPPs with 25~MHz
bandwidth at 430~MHz and 400~MHz at the higher frequencies. The GBT
data at 820~MHz were obtained using the Berkeley--Caltech Pulsar
Machine \citep{bdz+97} in the 48-MHz bandwidth mode described by
\cite{csl+02}. With the exception of 430~MHz, the integrated pulse
profiles (Figure~\ref{fig:mfprofs}) show that
the emission is characterized by a narrow main pulse and significant
interpulse feature. The exponential tails present at 430~MHz are most
likely due to interstellar scattering.  Flux density estimates based on
the off-pulse noise and the radiometer equation \citep[e.g.~][]{lk05} at
each of these frequencies are listed in Table~\ref{tab:parms}. Also listed
is the spectral index, $\alpha$, obtained 
from a fit to $S_\nu
\propto \nu^\alpha$, where $S_\nu$ is the flux density at an observing
frequency $\nu$.

To measure the polarization properties of PSR~J1906+0746, a 2.5-hr Parkes
observation at 1.4~GHz using a wideband correlator was carried out on
2005 May 25 which provided all four Stokes parameters across a 256-MHz
band.  Both the main pulse and interpulse have high linear
polarization (about 65\% and 50\% respectively) with position angle swings
of about $30^{\circ}$ across both pulses, increasing for the main
pulse and decreasing for the interpulse.  The observed frequency
dependence of position angle across the band implies a rotation
measure (RM) of $150 \pm 10$ rad~m$^{-2}$, consistent in
sign with the Faraday rotation seen toward most other pulsars and
extragalactic sources along nearby sightlines \citep{ccsk92}.

\section{Discussion}\label{sec:discussion}

\subsection{Nature of the companion}
\label{sec:nature}

From the orbital parameters of PSR~J1906+0746, we infer a Keplerian
mass function $f(m_p,m_c)=(m_c \sin i)^3/(m_p+m_c)^2 = 0.11 \,
M_{\odot}$. Here $m_p$ is the pulsar mass, $m_c$ is the companion mass
and $i$ is the angle between the orbital plane and the plane of the
sky.  Our measurement of the orbital periastron advance
$\dot{\omega}=7.57\pm0.03~$\arcdeg~yr$^{-1}$, is, after the double
pulsar system \cite{bdp+03}, the second largest observed so far.
Interpreting this large value within the framework of general
relativity implies that the total system mass
$M\,=\,m_p+m_c\,=\,2.61\pm0.02\,M_{\odot}$.  Using this, and the
constraints from the mass function, we infer the limits $m_c > 0.9 \,
M_{\odot}$ and $m_p < 1.7 \,\rm M_{\odot}$. Measured masses of
the neutron stars in double neutron star binary systems range from
1.25~M$_{\odot}$ \citep[for PSR~J0737$-$3039B,][]{lbk+04} to
1.44~M$_{\odot}$ \citep[for PSR~B1913+16,][]{wt03}, it is likely that
the mass of PSR~J1906+0746 is within these limits. If so, then
$1.17\,M_{\odot} < m_c < 1.36\,M_{\odot}$ and $42^\circ < i <
51^\circ$ respectively. These inclinations place the size of any
Shapiro delay well below the current level of timing precision.

Although it is possible, in principle, to account for the observed
$\dot{\omega}$ by classical effects \citep[e.g.~by a tidally induced
quadrupole moment from a main sequence star companion,][]{sb76}, as
discussed by \cite{klm+00a} these are unlikely for this particular
orbital configuration. The simplest interpretation is the relativistic
case and the above mass constraints suggest that the companion to
PSR~J1906+0746 is either a massive white dwarf or another neutron
star.  For the case of a white dwarf companion, the implication
\citep{dc87,py99,ts00a} would be that PSR~J1906+0746 formed from a
binary system of near unity mass ratio in which both stars were below
the critical core-collapse supernova mass limit $M_{\rm crit} \sim 8
M_{\odot}$. Following a phase in which the accretion of matter from
the evolved and more massive primary star onto the initially less
massive secondary pushed its mass above $M_{\rm crit}$, the secondary
underwent a supernova explosion to form the currently observable
pulsar.  Two probable endpoints of this scenario are the binary
pulsars B2303+46 \citep{vk99} and J1141--6545 \citep{klm+00a}.

The alternative case of a neutron star companion is particularly
attractive following the recent discovery of the double pulsar system
J0737--3039 \citep{bdp+03,lbk+04}. Given the small inferred
characteristic age and large magnetic field of PSR~J1906+0746, it
could be the young second-born neutron star. In this case, as for
PSR~J0737--3039A in the double pulsar system, we would expect the
companion to be a longer-lived recycled pulsar, spun up to a period of
a few tens of ms during an earlier phase of accretion.

To help establish the nature of the companion to PSR~J1906+0746, we
have carried out a search for radio pulsations in all data collected
so far. Following \cite{clm+04}, we have removed the deleterious
effects of the orbital motion on these observations by transforming
the time series (dedispersed to the DM given in Table~1) to the rest
frame of the companion.  To account for the unknown orbital
inclination, and hence mass ratio, of PSR~J1906+0746 this process was
carried out for 110 different assumed companion masses uniformly
sampled in the range $0.9 < m_c < 2.0~M_{\odot}$.  The corrected time
series were then searched for periodic signals by the same Fourier
transform analysis used in the original survey.  No convincing pulsar
candidates were seen down to S/N limits of 6. The most sensitive
observation we have made so far is a 2.2-hr Arecibo observation using
4 WAPPs to span a 400-MHz band centered at 1.4~GHz. Assuming a period
of 15~ms and a 10\% duty cycle for the putative pulsar, we estimate a
limiting flux density of 5$\mu$Jy.  At the nominal distance of
5.4~kpc, the corresponding 1.4-GHz luminosity limit is
0.1~mJy~kpc$^2$; only 0.5\% of all currently known pulsars
\citep{mhth05} have a luminosity below this value.  In parallel
to these deep periodicity searches, we have also carried out a search
for single pulses in the dedispersed time series. No significant
events were found down to flux density limits of 130~mJy at 1.4~GHz.

These deep searches suggest that the companion to PSR~J1906+0746 is
either: (a) a white dwarf; (b) a faint radio pulsar with a luminosity
below 0.1~mJy~kpc$^2$; or (c) a pulsar whose radio beam does not
intersect our line of sight. Option (a) will be hard to test
conclusively. A 100~kyr white dwarf at a distance of 5.5~kpc will be
of 24th magnitude. However, given the time necessary for the mass
transfer phase in the aforementioned evolutionary scenario, a more
probable age for the white dwarf might be $\sim 1$~Myr. For this
larger age, the expected visual magnitude is 29.  Even with the very
best optical observations it may therefore not be possible to rule out
this hypothesis. Given the sensitivity reached by our deep Arecibo
observations, option (b) is unlikely to be testable in the near
future. However, due to the high expected geodetic precession rate
(see below) it remains a possibility that the radio beam of the
putative companion pulsar will precess into our line of sight in
future. A further probe of the nature of the companion would be
orbital-phase dependent variations in the pulse profile of
PSR~J1906+0746. This effect is now well established in the double
pulsar binary system as the relativistic wind from PSR~J0737--3039A
impinges on the magnetosphere of its less energetic companion
J0737--3039B \citep{lbk+04}. While the larger orbital separation and
higher rate of spin-down in PSR~J1906+0746 will make any effect far
less for PSR~J1906+0746, it may be detectable in the future. There are
no significant pulse profile variations with orbital phase in our
current data.

\subsection{System age and implications for the compact-binary merger rate}\label{sec:age}

PSR~J1906+0746 has the smallest characteristic age ($\tau_c=112$~kyr)
of any binary pulsar currently known. It is also worth noting that, as
seen for other young pulsars \citep[e.g.][]{vkk98}, a significant
amount of polarization is observed.  As supernova remnant associations
are known for isolated pulsars with similar spin characteristics as
PSR~J1906+0746 \citep[e.g.~PSR~B0656+14,][]{tbb+03}, we have searched
the available literature for signs of diffuse radio emission. No
candidates are present in the most recent compilation of supernova
remnants \citep{gre04}.  Similarly, no extended radio emission is
present at or near the pulsar position in a 332-MHz VLA observation of
this region \citep{kkfv02}.  The corresponding 1-$\sigma$ surface
brightness upper limit of
$\sim6\times10^{-22}$~W~m$^{-2}$~Hz$^{-1}$~sr$^{-1}$ which is fainter
than almost all known supernova remnants \citep{gre04}.

Regardless of the exact nature of the companion to PSR~J1906+0746, the
small $\tau_c$ for this pulsar is unusual when placed in context of
the radio lifetimes of 10--100~Myr estimated for normal pulsars
\cite[see e.g.][]{lmt85,bwhv92}.  Unless $\tau_c$ is a significant
underestimate of the true age of this pulsar, the implication is that
we have found a pulsar which is only 0.1--1\% through its active radio
lifetime.  Since this is highly unlikely to occur by chance, the youth
of PSR~J1906+0746 suggests a potentially significant birth rate for
this new class of objects.

\begin{figure}
\includegraphics[scale=0.7,angle=270]{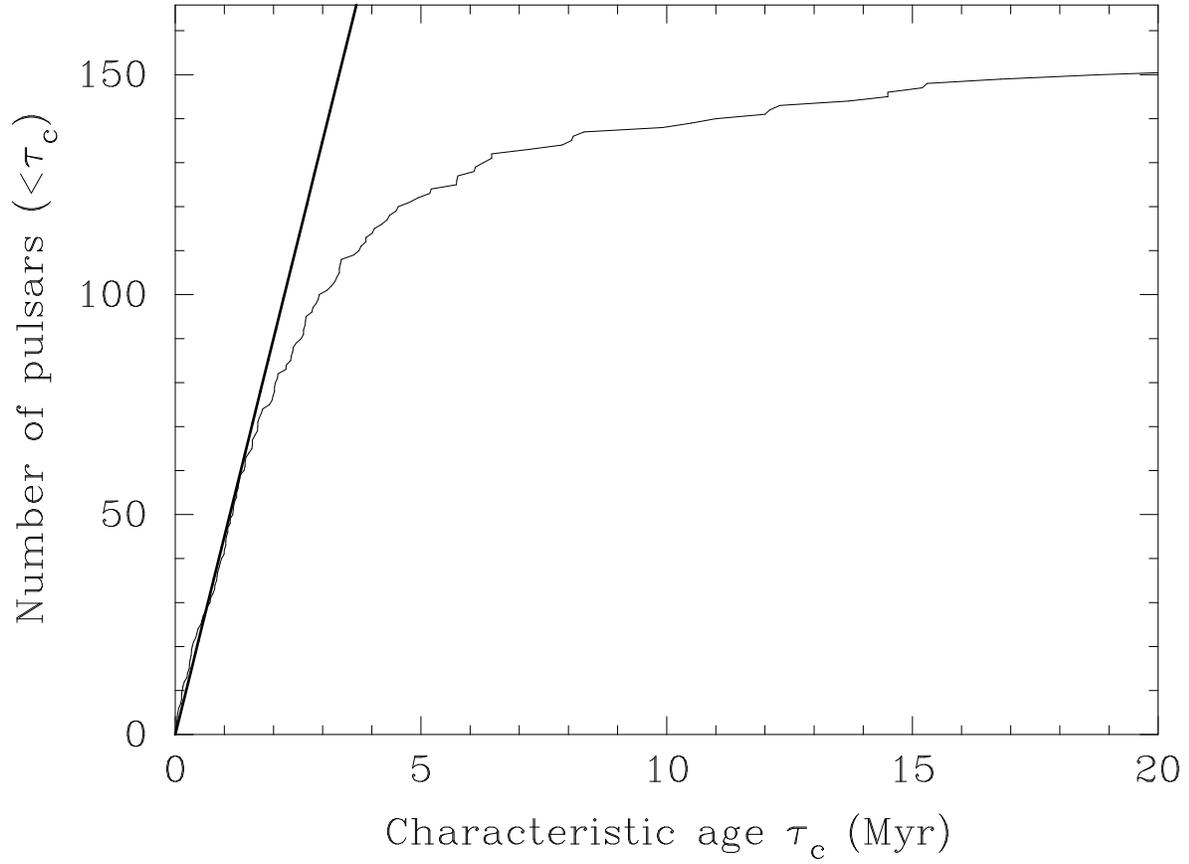}
\caption{\label{fig:cage} 
Cumulative distribution of characteristic ages for pulsars detected
in the PMPS with similar surface magnetic fields to PSR~J1906+0746. The
heavy solid line shows the expected trend corresponding to a birthrate
of potentially observable objects of 45~Myr$^{-1}$ (see text).}
\end{figure}

A simple estimate of the birth rate can be made by considering the
cumulative distribution of characteristic ages. Figure~\ref{fig:cage}
shows this distribution for a sub-sample of 150 pulsars detected in
the PMPS which have inferred magnetic field strengths within 1~dB of
PSR~J1906+0746 (i.e.~$|\log(B)-\log(B_{1906})|<0.1$). This sample
shows a linear trend at small ages with a slope of 45~Myr$^{-1}$. The
relative dearth of pulsars with larger ages reflects the difficulty of
detecting them (e.g.~due to luminosity decay, period-dependent beaming
effects, or motion away from the Galactic disk). Nevertheless, the
slope of the trend does provide a good estimate of the formation rate
of normal pulsars detectable by the PMPS with similar magnetic fields
to PSR~J1906+0746. Since the PMPS has only detected one
J1906+0746-like pulsar in this sample, the inferred birthrate of {\it
similar pulsars potentially observable in the PMPS} is
$45/150=0.3$~Myr$^{-1}$.

To scale this rate over the whole Galaxy, we used a Monte Carlo
simulation to estimate the total Galactic volume probed by the PMPS
for J1906+0746-like objects. The simulation seeded a model galaxy with
pulsars following the distribution and luminosity functions derived by
\cite{lor04}. We find that the PMPS should detect roughly one
J1906+0746-like pulsar for every 40 in the Galaxy. With this scaling,
we estimate the birth rate of PSR~J1906+0746-like objects to be $\sim
60 \times (0.2/f)$~Myr$^{-1}$, where $f$ is the unknown fraction of
$4\pi$~sr covered by the radio beam. For a steady-state population of
compact binaries similar to PSR~J1906+0746, we can reasonably assume
that the merging rate is equal to the birth rate. Our first-order
estimate therefore implies a merger rate of J1906+0746-like binaries
similar to that derived recently by \citep{kkl+04} for double neutron
star binaries.  Once the nature of the companion is established, the
inclusion of PSR~J1906+0746 in future population synthesis
calculations will provide an important additional constraint on the
merger rate of compact binaries (Kim et al.~2003, 2004).

The orbital eccentricity of 0.085 for PSR~J1906+0746 is
remarkably similar to that of PSR~J0737$-$3039 \citep{bdp+03}, but
smaller than observed for other double neutron star systems. As
proposed recently by \cite{cb05}, for PSR~J0737$-$3039 this is likely
a selection effect: systems with similar orbital periods but
significantly higher eccentricities would coalesce on much shorter
timescales than the likely age of PSR~J0737$-$3039 ($\sim
10^8$~yr). As a result, more eccentric binaries are much less likely
to be observed. For systems as young as PSR~J1906+0746, where there
has been less time for significant gravitational wave decay, systems
with higher eccentricities could in principle be observed. Until
further examples of young compact binary pulsars are discovered, it is
difficult to conclude whether the low eccentricity is a necessary
feature of these systems. If it is, we will have better constraints on
the sequence of events that leads to the formation of pulsars,
including supernova kicks \citep[see e.g.~][]{dpp05}.  If it is not,
then there is the exciting possibility of finding other young pulsars
in even more compact and eccentric orbits.

\subsection{Pulse profile evolution}\label{sec:profevol}

It is now established for pulsars observed in four other relativistic
binaries (PSR~B1913+16: Weisberg et al.~1989; Kramer 1998:
Weisberg \& Taylor 2002; PSR~B1534+12: Stairs et al.~2004; 
PSR~J1141$-$6545: Hotan et al.~2005; PSR~J0737$-$3039B: Burgay
et al.~2005) that the mean pulse profile varies with time. With the
possible exception of PSR~J0737--3039B \citep{bpm+05}, the simplest
explanation for this effect is geodetic precession --- a general
relativistic effect in which the pulsar spin axis precesses about the
total system angular momentum \citep{dr74}. The profile variations
occur as the precessing pulsar beam changes its orientation with
respect to our line of sight. Given the highly relativistic nature of
PSR~J1906+0746, we also expect to observe this effect in this system.
Using the above mass constraints, and the standard geodetic precession
formulae \citep{bo75}, for PSR~J1906+0746 we find the expected
precession rate is 2.2~\arcdeg~yr$^{-1}$ for the double neutron star
case, or 1.6~\arcdeg~yr$^{-1}$ for the neutron star--white dwarf
case. These correspond respectively to precession periods of 164 and
225 yr. The change in the angle between the line-of-sight and the
spin axis is likely to be significantly smaller than the precession
rates imply themselves \cite[see e.g.,][]{hbo05}.

\begin{figure}
\includegraphics[scale=0.6,angle=270]{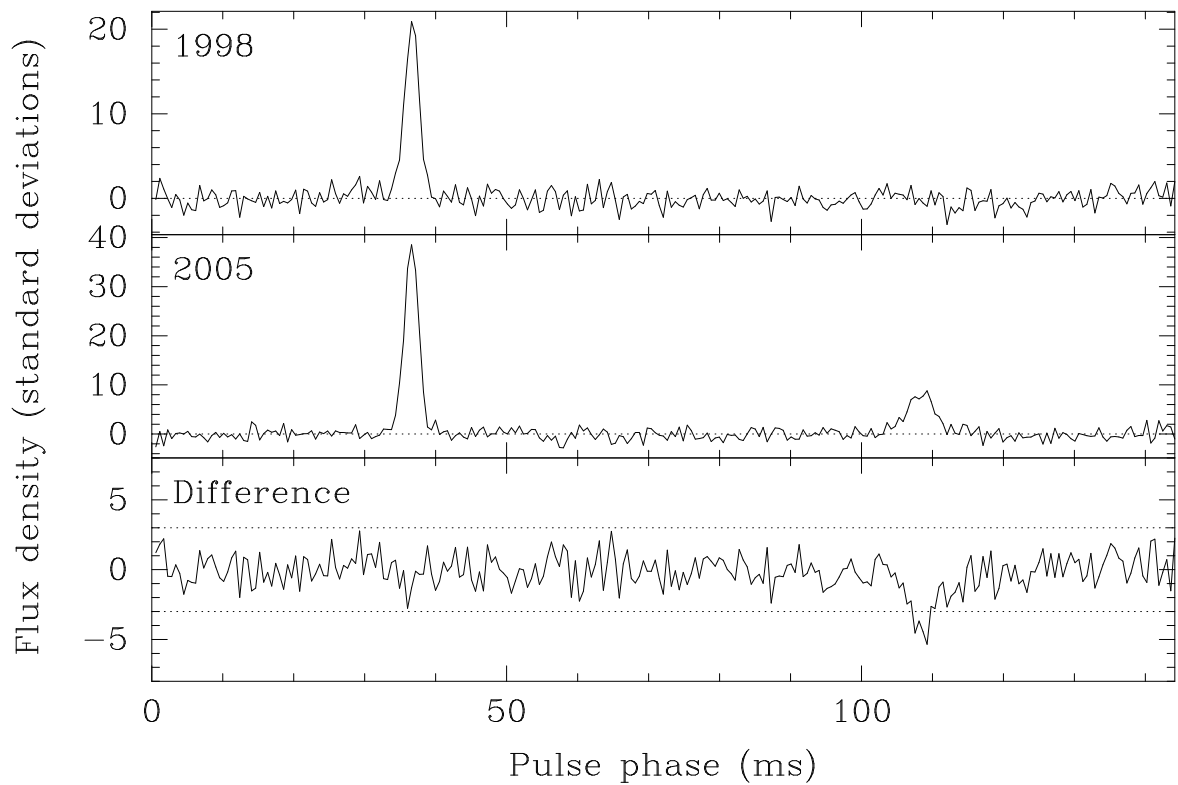}
\caption{\label{fig:profs} Integrated pulse profiles of PSR~J1906+0746
showing $360^\circ$ of rotational phase. The upper panel shows the
detection at 1.374~GHz from the 35-min of PMPS data taken on 1998
August 3. The lower panel shows a 35-min observation with the same
observing system taken on 2005 September 4.  The lower panel shows
the difference profile (i.e.~1998 minus 2005 data) after scaling both
profiles to the area of the main pulse. The dashed horizontal lines
show $\pm 3$ standard deviations computed from the off-pulse noise region.
The limiting instrumental time resolution of both these profiles is 2.1~ms.}
\end{figure}

The PMPS detection of PSR~J1906+0746 provides a 7-yr baseline to look
for pulse-shape changes in this new system.  Figure~\ref{fig:profs}
shows a comparison between the integrated profiles at 1400~MHz from
the 1998 PMPS detection and a recent Parkes observation at the same
frequency using the same observing system.  We have scaled each
profile to the area of the main pulse and formed the ``difference
profile'' by subtracting the 2005 profile from the 1998 one. In the
absence of any profile evolution, the difference profile should be
free from systematic trends and have a standard deviation consistent
with the quadratic sum of the off-pulse noise present in the two input
profiles. We observe a significant departure from random noise around
the interpulse region which is not detectable in the 1998 observation.
It is also notable that the peak S/N (which we define as the peak
of the main pulse divided by the off-pulse standard deviation) seen
in 2005 is 80\% larger than observed in 1998.  Accounting for
variations in telescope gain for the two observations, and the fact
that the 1998 observation was offset from the nominal position by 3.8
arcmin, we expect a S/N increase of only 30\%. As scintillation is not
expected at this frequency for this DM, these observations suggest
that the flux density of PSR~J1906+0746 has increased since
1998. Further observations with better S/N, flux calibration and
polarimetric capability are required to confirm and quantify these
changes.

\section{Conclusions and future work}\label{sec:future}

We have presented the discovery and initial follow-up observations of
PSR~J1906+0746, a young 144-ms pulsar in a highly relativistic 3.98-hr
orbit about a $>0.9~M_{\odot}$ companion. The new system emphasizes
the relative immunity of surveys with short integration times to
the smearing of pulsar signals caused by fast orbital motion. It also
shows the value of storing archival search data for retrospective
analysis. The orbital characteristics and total mass of the binary
system, $2.61\pm0.02~M_{\odot}$, inferred from the measured periastron
advance $\dot{\omega}=7.57\pm0.03$~\arcdeg~yr$^{-1}$, suggest that the
companion is either another neutron star or a massive white dwarf. In
the former case, the 112-kyr characteristic age of PSR~J1906+0746
would imply that its companion is a short-period recycled neutron
star. However, despite intensive searches, we have been so far unable
to detect radio pulsations from the companion. Optical searches for a
white-dwarf companion could possibly clarify the situation.
Regardless of the nature of the companion, the apparent youth of
PSR~J1906+0746 implies a birth rate of $\sim 60$~Myr$^{-1}$ for
similar objects in the Galaxy. Note that this estimate assumes a
beaming fraction for these objects of 20\%.

\begin{figure}
\includegraphics[scale=0.9,angle=270]{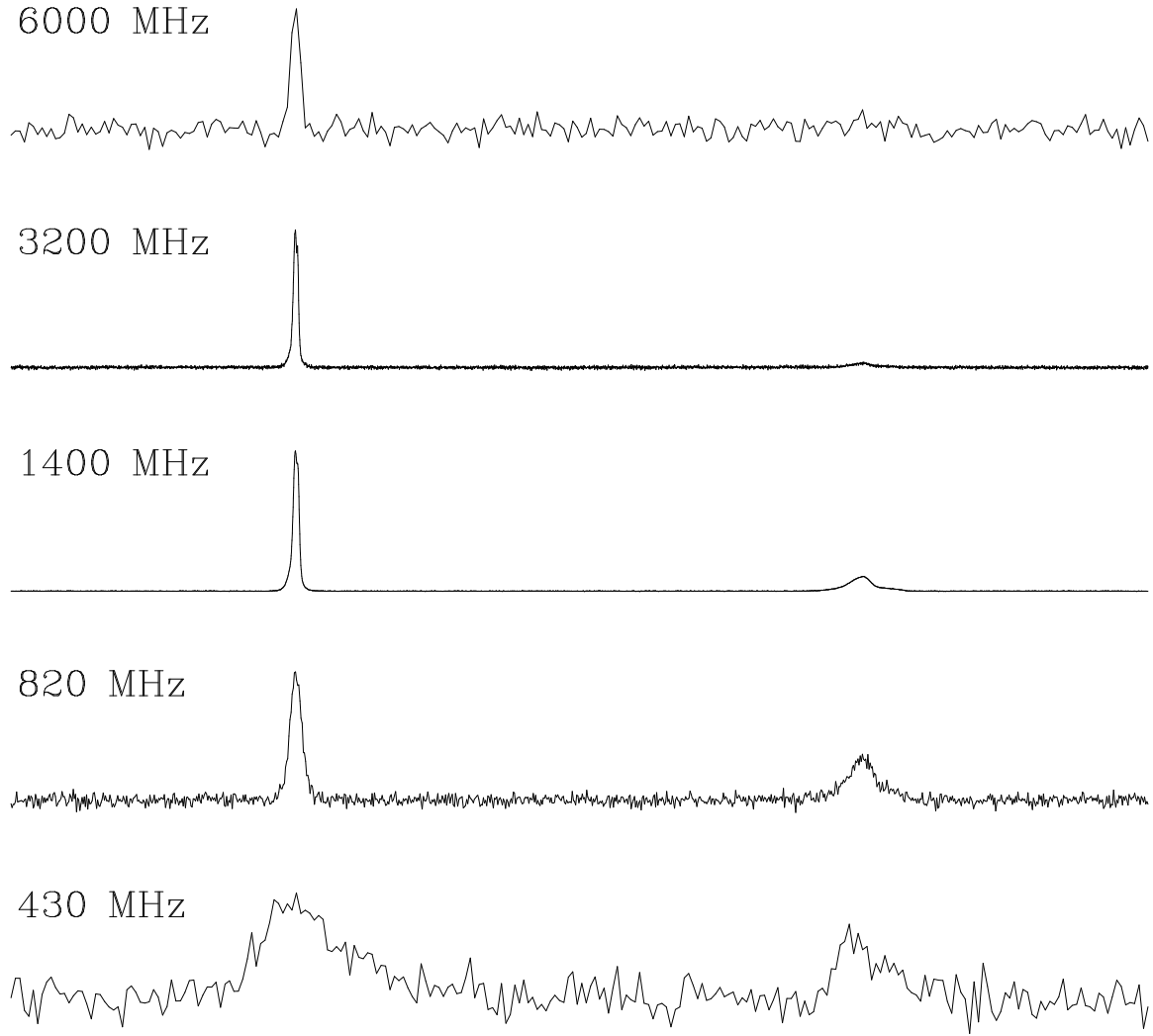}
\caption{\label{fig:mfprofs}
Multi-frequency pulse profiles for PSR~J1906+0746 obtained from observations
at Arecibo (430~MHz, 1400~MHz, 3200~MHz and 6000~MHz) and Green Bank
(820~MHz). Each profile shows 360\arcdeg of rotational and is freely
available on-line as part of the European Pulse Network database
(http://www.jb.man.ac.uk/$\sim$pulsar/Resources/epn).
}
\end{figure}

Future radio timing and polarimetric observations of PSR~J1906+0746
should allow the study of several relativistic effects. The very
narrow pulse shape observed (Figure~\ref{fig:mfprofs}) means that our
current Arecibo TOAs have an uncertainty of $\sim 5\mu$s in a 5-min
integration. Simulations based on this level of precision show that we
expect to measure the gravitational redshift and time dilation
parameter, $\gamma$, within the coming year and, within a few years,
measure the rate of orbital decay, $\dot{P}_b$. Such measurements
would determine the orbital inclination and masses of the stars so
that the system could possibly be used for further tests of general
relativity. Assuming reasonable ranges for the pulsar mass, the
orbital inclination angle is likely to be in the range
$42^{\circ}<i<51^{\circ}$. For this range of inclinations, a
measurement of the Shapiro delay is less likely at the present level
of precision.  The distance to PSR~J1906+0746 is currently estimated
using the \cite{cl02a} electron density model. As for PSR~B1913+16,
kinematic contributions to $\dot{P}_b$, which depend on the assumed
location in the Galaxy and hence the distance, are likely to be a
limiting factor for high-precision tests of general relativity with
this system. We are currently attempting to obtain independent
distance constraints via the detection of neutral hydrogen absorption
and emission.

Comparing pulse profiles taken in 1998 and 2005, we have found some
evidence for long-term evolution of the pulse profile.  Given that the
expected geodetic precession period is only $\sim 200$~yr, the
observed variations could be the first manifestations of this
effect. Future observations with high time resolution and polarimetric
capabilities should provide more quantitative insights and may permit
a mapping of the radio beam of PSR~J1906+0746, as has been possible
for PSR~B1913+16 \citep{wt02}.

\acknowledgements

We thank the staff at NAIC and ATNF for developing ALFA and its
associated data acquisition systems.  In particular, we thank Arun
Venkataraman, Jeff Hagen, Bill Sisk and Steve Torchinsky at NAIC and
Graham Carrad at the ATNF. This work was supported by the NSF through
a cooperative agreement with Cornell University to operate the Arecibo
Observatory.  NSF also supported this research through grants
AST-02-05853 and AST-05-07376 (Columbia University), AST-02-06035 and
AST-05-07747 (Cornell University) and AST-02-06205 (Princeton
University). D.R.L is a University Research Fellow funded by the Royal
Society.  I.H.S holds an NSERC UFA, and pulsar research at UBC is
supported by a Discovery Grant. Z.A acknowledges support from grant
NRA-99-01-LTSA-070 to NASA GSFC. L.K holds an NSERC Canada Graduate
Scholarship.  N.D., A.P., and M.B. received support from the Italian
Ministry of University and Research under the national program Cofin
2003.  The Parkes telescope is part of the Australia Telescope, which
is funded by the Commonwealth Government for operation as a National
Facility managed by CSIRO.  NRAO is a facility of the NSF operated
under cooperative agreement by Associated Universities, Inc.  Research
in radio astronomy at the NRL is supported by the Office of Naval
Research.  We acknowledge useful discussions with Cees Bassa
concerning the optical follow-up of the putative white dwarf
companion. Finally, we wish to thank the referee, Matthew Bailes, for
useful comments on an earlier version of this manuscript.

\begin{deluxetable}{ll}
\tablewidth{0pt}
\tablecaption{\label{tab:parms}Observed and derived parameters of 
PSR~J1906+0746.}
\tablecomments{
Figures in parentheses are 1-$\sigma$ uncertainties in the
least-significant digit(s) and the constant $\rm T_{\odot} \equiv
GM_{\odot}$$/c^3 \simeq 4.925 \mu$s.  Definitions for $B$, $\tau_c$ and
$\dot{E}$ are from \cite{lk05}. The distance estimate uses the pulsar
position, DM and the Cordes \& Lazio (2002) Galactic electron density model.
}
\tablecolumns{2}
\tablehead{
\colhead{Parameter} &
\colhead{Value}
}
\startdata  
Right ascension (J2000)\dotfill&$19^{\rm h}06^{\rm m}48\fs673(6)$\\
Declination (J2000)\dotfill    & $07\arcdeg46\arcmin28.6(3)\arcsec$ \\
Spin period, $P$ (ms)\dotfill                   & 144.071929982(3)        \\
Spin period derivative, $\dot P$\dotfill      & $2.0280(2)\times 10^{-14}$  \\
Epoch (MJD)\dotfill                                 & 53590               \\
Orbital period, $P_b$ (days)\dotfill                & 0.165993045(8)\\
Projected semi-major axis, $x$ (lt s)\dotfill          & 1.420198(2)\\
Orbital eccentricity, $e$\dotfill                   & 0.085303(2)\\
Epoch of periastron, $T_0$ (MJD)\dotfill            & 53553.9126685(6)\\
Longitude of periastron, $\omega$ (deg)\dotfill     & 61.053(1)\\
Periastron advance rate, $\dot{\omega}$ (\arcdeg~yr$^{-1}$)\dotfill&7.57(3)\\
Dispersion measure, DM (cm$^{-3}$\,pc)\dotfill      & 217.780(2)          \\
Rotation measure, RM (rad m$^{-2}$)\dotfill            & +150(10) \\
Flux density at 0.4\,GHz, $S_{0.4}$ (mJy)\dotfill & 0.9(2)   \\
Flux density at 0.8\,GHz, $S_{0.8}$ (mJy)\dotfill & 0.72(15)   \\
Flux density at 1.4\,GHz, $S_{1.4}$ (mJy)\dotfill & 0.55(15)   \\
Flux density at 3.2\,GHz, $S_{3.2}$ (mJy)\dotfill & 0.12(3)\\
Flux density at 6.0\,GHz, $S_{6.0}$ (mJy)\dotfill & 0.030(7)\\
Main pulse widths at 50 and 10\% (ms)\dotfill& 0.6 and 1.7\\
Characteristic age, $\tau_c=P/2\dot{P}$ (kyr)\dotfill     & 112             \\
Magnetic field, $B=3.2 \times 10^{19} (P \dot{P})^{1/2}$
 (Gauss)\dotfill         & $1.7\times 10^{12}$ \\
Spin-down power, $\dot E = 3.95 \times 10^{46}
\dot{P}/P^3$ (ergs\,s$^{-1}$)\dotfill& $2.7\times10^{35}$\\
Inferred distance, $d$ (kpc)\dotfill           & $\sim 5.4$        \\
Spectral index, $\alpha$\dotfill & --1.3(2)\\
Radio luminosity at 1.4\,GHz, $S_{1.4}d^2$ (mJy\,kpc$^2$)\dotfill &$\sim16$\\
Mass function, $f(m_p,m_c)=4\pi^2x^3/(\rm T_{\odot}$$P_b^2)$
($M_{\odot}$)\dotfill  &  0.1116222(6)\\
Total system mass, $M=m_p+m_c$, ($M_{\odot}$)\dotfill  &  2.61(2)\\
Gravitational wave coalescence time, $\tau_g$ (Myr)\dotfill& $\sim 300$
\enddata
\end{deluxetable}
\end{document}